\documentclass[aps,prl,twocolumn,showpacs]{revtex4}
\usepackage{bm}
\usepackage{epsfig}
\usepackage{graphicx}
\usepackage{amssymb,amsmath,amsbsy,amsgen,amsfonts}    
\usepackage{dcolumn}
\usepackage{amsthm}
\usepackage{mathrsfs}
\usepackage{latexsym}
\usepackage{array}
\usepackage{amstext}

\usepackage{epstopdf} 

\newcommand{\ket}[1]{\left\vert{#1}\right\rangle}
\newcommand{\ketbra}[2]{|#1\rangle \langle#2|}
\newcommand{\be}{\begin{equation}}
\newcommand{\ee}{\end{equation}}
\newcommand{\ba}{\begin{array}}
\newcommand{\ea}{\end{array}}
\newcommand{\bqa}{\begin{eqnarray}}
\newcommand{\eqa}{\end{eqnarray}}

\begin{document}

\title{A scalable method for demonstrating the Deutsch-Jozsa \\ and Bernstein-Vazirani algorithms using cluster states} 

\author{M. S. Tame and M. S. Kim} 
\affiliation{$^1$Institute for Mathematical Sciences, Imperial College London, SW7 2PG, United Kingdom \\ $^2$QOLS,\,The\,Blackett\,Laboratory,\,Imperial\,College\,London,\,Prince\,Consort\,Road,\,SW7\,2BW,\,United Kingdom}
\date{\today}

\begin{abstract}
We show that fundamental versions of the Deutsch-Jozsa and Bernstein-Vazirani quantum algorithms can be performed using a small entangled cluster state resource of only six qubits. We then investigate the minimal resource states needed to demonstrate general $n$-qubit versions and a scalable method to produce them. For this we propose a versatile on-chip photonic waveguide setup. 
\end{abstract}

\pacs{03.67.-a, 03.67.Mn, 42.50.Dv, 03.67.Lx}

\maketitle

In recent years considerable progress has been made in the realization of {\it quantum technology} based on a wide-range of physical settings~\cite{OBrien}. Most notably the demonstration of simple logic gates for quantum computing (QC) has been achieved. However, piecing these logic gates together in order to perform quantum algorithms capable of outperforming their classical analogues~\cite{deutsch, DJ, BV, algoritmi} is still far from being practical. An approach to QC called the {\it one-way} model~\cite{oneway} promises to help overcome this major problem. Here, a multipartite entangled state, the {\it cluster state}, is used as a resource to perform QC where the amount of control one needs over a quantum system is reduced to the ability of performing just single-qubit measurements. This is an important advantage for a variety of physical systems, such as those using photons, where already the model's general features have been experimentally demonstrated in setups with bulk components~\cite{onewayexp1,onewayexp2,onewayexp3}. We are now faced with two key challenges. The first is to identify the minimal resources needed for performing quantum protocols so that small-scale algorithms can be demonstrated and eventually scaled up. Second, it is vital that optimal methods are found for transferring what has been done in bulk setups to more practical and ultimately scalable settings~\cite{OBrien2}. 

In this work we address these two important challenges. We show that fundamental versions of the Deutsch-Jozsa (DJ)~\cite{DJ} and Bernstein-Vazirani (BV)~\cite{BV} algorithms can be demonstrated using one-way QC on a small entangled cluster state of only six qubits. Moreover, we investigate the resources needed for demonstrating general $n$-qubit versions of the algorithms. We then propose a versatile photonic on-chip setup where they could be realized and eventually scaled-up. Our proposal represents the first scalable method for demonstrating these algorithms in the promising context of one-way QC. Recently on-chip and waveguide setups have started to gain momentum as prominent settings for full-scale QC. Here different components - requiring expertise from a broad range of the physical sciences - must be integrated together, making this a truly multidisciplinary endeavor. However, once a given setup is fabricated it is not possible to change it. Our proposal addresses this problem by providing an architecture of minimal resources on which to realize various algorithms. 

{\it DJ Algorithm.-} We start by briefly reviewing the DJ algorithm~\cite{DJ} which takes an $n$-bit binary input $x \in \{0,1\}^{n}$ and allows one to distinguish two types of function $f(x)$ that apply the transformation $f(x): \{0,...,2^{n}-1\} \mapsto \{0,1\}$ implemented by an oracle. $f(x)$ is {\it constant} if it returns the same value (0 or 1) for all inputs and {\it balanced} if it returns 0 for half the inputs and 1 for the other half. Classically the oracle must be queried as many as $(2^{n}/2)+1$ times in the worst case. Quantum mechanically only one query is required in all cases. 

The quantum algorithm begins with the state $2^{-n/2}\sum_{x\in \{0,1 \}^{n}}\ket{x}\ket{y}=\ket{+}^{\otimes n}\ket{-}$, where $\ket{y}$ is an ancilla qubit, $\ket{\pm}=(\ket{0}\pm \ket{1})/\sqrt{2}$ and $\{ \ket{0},\ket{1}\}$ is the single-qubit computational basis. The oracle applies the transformation $\ket{x}\ket{y}\mapsto \ket{x}\ket{y\oplus f(x)}$, producing the state $2^{-n/2}\sum_{x \in \{0,1 \}^{n}}(-1)^{f(x)}\ket{x} \ket{-}$. Hadamard gates ${\sf H}=(\sigma_x + \sigma_z)/\sqrt{2}$ are then applied to all the query qubits ($\sigma_{x,y,z}$ are the Pauli matrices), resulting in the state $\ket{\psi_{out}}=2^{-n}\sum_{z}\sum_{x} (-1)^{x \cdot z + f(x)} \ket{z}\ket{-}$ (where $x \cdot z$ is the bitwise inner-product of $x$ and $z$). The amplitude for the state of the query qubits as $\ket{0}^{\otimes n}$ is $\sum_{x}[(-1)^{f(x)}/2^n]$. There are two cases: First, if $f(x)$ is {\it constant} then the amplitude for $\ket{0}^{\otimes n}$ is +1 or -1 depending on the constant value $f(x)$ takes. As $\ket{\psi_{out}}$ is of unit length, all other amplitudes go to zero. Second, if $f(x)$ is {\it balanced}, then the positive and negative contributions to $\ket{0}^{\otimes n}$ cancel, leaving an amplitude of zero. Therefore if $\ket{\psi_{out}}=\ket{0}^{\otimes n}$ is measured, $f(x)$ is constant, otherwise $f(x)$ is balanced. Thus, only one query is required. 
A {\it refined} version of the above algorithm has been suggested~\cite{refDJ}. Here, one identifies that the ancilla qubit $\ket{y}$ is unentangled both before and after the black box operation and can be removed~\cite{cl}. The oracle's action then reduces to $\ket{x}\mapsto (-1)^{f(x)}\ket{x}$, with the remainder of the algorithm performed as described above. For a one-qubit query register (DJ$_1$), the algorithm corresponds to Deutsch's problem~\cite{DJ}, for which it is known a four-qubit cluster state can be used~\cite{onewayexp2}. Thus in what follows we consider the fundamental case of DJ$_2$. We then discuss the corresponding refined version and finally arbitrary $n$-qubit versions. 
\begin{figure}[t]
\psfig{figure=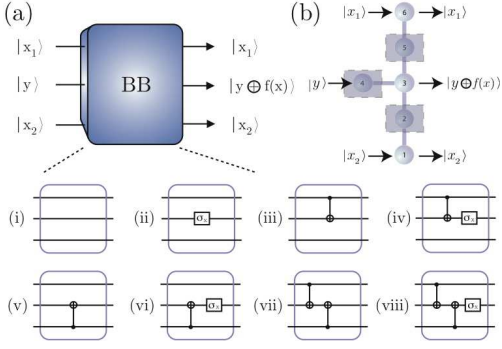,width=6.6cm}
\caption{Black box (BB) circuits for the Deutsch-Jozsa algorithm and cluster state. {\bf (a)}~BB's (i) and (ii) ((iii)-(viii)) correspond to constant (balanced) functions. See Table \ref{tab1} for $f(x)$ in each case. {\bf (b)}: The six-qubit cluster state $\ket{\Phi_{C}}$ depicting the input/output logical qubits. The measurements and outcomes of qubits $1$, $3$ and $6$ constitute the algorithm. The measurements of qubits 2, 4 and 5 (and their feed-forward) are carried out by the oracle, thus boxes surround these qubits.}
\label{blackbox}
\end{figure}

{\it Cluster state implementation.-} The action of an oracle is preset or dictated by the outcome of another algorithm and is known as a {\it promise} problem~\cite{DJ}. In order to implement all the functions that it might use for DJ$_2$, we must be able to construct them using a combination of quantum gates. 
In Fig.~\ref{blackbox}~{\bf (a)} we show all the oracle functions in terms of their quantum network. By describing each as a ``black box'', one can see that all eight black boxes (BB(i)-(viii)) implement their respective oracle function given in Table \ref{tab1}. In order to carry out the algorithm using these quantum gates, the six-qubit cluster state resource shown in Fig.~\ref{blackbox}~{\bf (b)} can be used, where one-way QC is carried out by performing a program of measurements. No adjustment to the resource is necessary; each BB corresponds to a different measurement program on the same resource.

For a cluster state two types of single-qubit measurements allow one-way QC to be performed. First, measuring a qubit $j$ in the computational basis removes it from the cluster, leaving a smaller cluster state of the remaining qubits. Second, in order to perform QIP, qubits must be measured in the basis $B_j(\alpha)=\{ \ket{\alpha_+}_j,\ket{\alpha_-}_j \}$, where $\ket{\alpha_{\pm}}_j=(\ket{0}\pm e^{i \alpha}\ket{1})_j/\sqrt{2}$ ($\alpha\!\in\!{\mathbb R}$). Choosing the measurement basis determines the rotation $R_z(\alpha)={\rm exp}(-i \alpha \sigma_z/2)$, followed by a Hadamard operation ${\sf H}$ applied to an encoded logical qubit in the cluster residing on qubit $j$. With a suitable cluster, any quantum logic operation can be performed using a measurement program ${\cal M}$ with an appropriate choice for the $B_j(\alpha)$'s~\cite{clusterback}.

The cluster state resource of Fig.~\ref{blackbox}~{\bf (b)} is given by $\ket{\Phi_C}=\frac{1}{2\sqrt{2}}[\ket{\phi_2^+}_{12}\ket{0+}_{34}\ket{\phi_2^+}_{65}+\ket{\phi_2^-}_{12}\ket{1-}_{34}\ket{\phi_2^-}_{65}]$,
where $\ket{\phi_2^\pm}_{ab}=(1/\sqrt{2})(\ket{+}\ket{0} \pm \ket{-}\ket{1})$. A set of measurement bases for the qubits in this resource can be used for each BB. In Table~\ref{tab1} we provide these bases and feed-forward (FF) operations~\cite{onewayexp1,onewayexp2,onewayexp3}. Fig.~\ref{blackbox}~{\bf (b)} shows the in-out states of the algorithm, where the logical input corresponding to $\ket{x}=\ket{x_1}\ket{x_2}=\ket{+}\ket{+}$ is encoded on qubits 1 and 6. The state $\ket{y}=\ket{-}$ is encoded on qubit 4 using a $\sigma_z$ operation before the measurement program begins: the resource $\ket{\Phi_C'}=(\sigma_z)_4\ket{\Phi_C}$ remains as a cluster state with the state $\ket{x_1}\ket{x_2}\ket{y}\equiv\ket{+}\ket{+}\ket{-}$ residing on the logical input register. 
\begin{table}[b]
\begin{ruledtabular}
\begin{tabular}{|c|c|c||c|c|c|c|c|c|}
\hline
\phantom{p}$f(x)$ & $i$ & $ii$ & $iii$ & $iv$ & $v$ & $vi$ & $vii$ & $viii$ \\ \hline
\phantom{p}$f(0)$ & 0 & 1 & 0 & 1 & 0 & 1 & 0 & 1 \\
\phantom{p}$f(1)$ & 0 & 1 & 0 & 1 & 1 & 0 & 1 & 0 \\
\phantom{p}$f(2)$ & 0 & 1 & 1 & 0 & 0 & 1 & 1 & 0 \\
\phantom{p}$f(3)$ & 0 & 1 & 1 & 0 & 1 & 0 & 0 & 1 \\
\hline \hline
${\cal M}_2$ & $\ket{0/1}$ & $\ket{0/1}$ & $\ket{0/1}$ & $\ket{0/1}$ & $B(\frac{\pi}{2})$ & $B(\frac{\pi}{2})$ & $B(\frac{\pi}{2})$ & $B(\frac{\pi}{2})$ \\
${\cal M}_4$ & $B(0)$ & $B(0)$ & $B(0)$ & $B(0)$ & $B(0)$ & $B(0)$ & $B(0)$ & $B(0)$ \\
${\cal M}_5$ & $\ket{0/1}$ & $\ket{0/1}$ & $B(\frac{\pi}{2})$ & $B(\frac{\pi}{2})$ & $\ket{0/1}$ & $\ket{0/1}$ & $B(\frac{\pi}{2})$ & $B(\frac{\pi}{2})$ \\
\hline \hline
FF$_1$ & ${\sf H}\sigma_z^{s_2}$ & ${\sf H}\sigma_z^{s_2}$ & ${\sf H}\sigma_z^{s_2}$ & ${\sf H}\sigma_z^{s_2}$ & ${\sf H}\chi^{s_2}$ & ${\sf H}\chi^{s_2}$ & ${\sf H}\chi^{s_2}$ & ${\sf H}\chi^{s_2}$ \\
FF$_3$ & $\zeta$ & $\sigma_x \zeta$ & $\tilde{\zeta}$ & $\sigma_x \tilde{\zeta}$ & $\tilde{\zeta}$ & $\sigma_x \tilde{\zeta}$ & $\zeta \sigma_z$ & $\sigma_x \zeta \sigma_z$ \\
FF$_6$ & ${\sf H}\sigma_z^{s_5}$ & ${\sf H}\sigma_z^{s_5}$ & ${\sf H}\chi^{s_5}$ & ${\sf H}\chi^{s_5}$ & ${\sf H}\sigma_z^{s_5}$ & ${\sf H}\sigma_z^{s_5}$ & ${\sf H}\sigma_z^{s_5}$ & ${\sf H}\sigma_z^{s_5}$ \\
\hline
\end{tabular}
\end{ruledtabular}
\caption{Black box outputs for the DJ algorithm and measurement program ${\cal M}_i$ for qubit $i$ of the cluster state with feed-forward operations. The notation $\ket{0/1}$ corresponds to a measurement in the computational basis, $\chi^{i}=\sigma_z^{i+s_4}R_z(-\pi/2)$, $\zeta= \sigma_z^{s_4}\sigma_x^{s_2+s_5}{\sf H}$ and $\tilde{\zeta}= \zeta R_z(-\pi/2)$. The value $s_j$ corresponds to the outcome of qubit $j$'s measurement.}
\label{tab1}
\end{table}

Qubits 2 and 5 in $\ket{\Phi_{C}'}$ play the pivotal role of the oracle by performing two-qubit gates between the logical input states $\ket{x_1}$, $\ket{x_2}$ and $\ket{y}$. For BB(i), measuring qubits 2 and 5 in the computational basis disentangles them from the cluster and $\ket{\Phi_{C}'}$ is transformed into $(1/2 \sqrt{2})(\ket{0}+(-1)^{s_2}\ket{1})_1(\ket{0}\ket{-}+(-1)^{s_2 \oplus s_5}\ket{1}\ket{+})_{34}(\ket{0}+(-1)^{s_5}\ket{1})_6$ (where $s_i$ corresponds to the outcome of qubit $i$'s measurement). 
The logical operation performed by this choice of the oracle's measurement basis is $\openone \otimes \openone  \otimes \openone$ (up to FF operations). By including the ${\sf H}$ operation applied to the state $\ket{y}$ from the measurement of qubit 4 in the basis $B_4(0)$, the overall computation results in $\openone \otimes \openone \otimes {\sf H} \ket{+}\ket{+}\ket{-}$ which is equivalent to $\ket{x}\ket{y \oplus f(x)}=(\openone \otimes \openone \otimes \openone)\ket{+}\ket{+}\ket{-}$ up to a local rotation ${\sf H}$ on qubit 3 incorporated into the FF stage. Qubits 1, 3 and 6 can now be taken as the output state $\ket{x}\ket{y \oplus f(x)}$. On the other hand, for BB(iii), upon measuring qubit 5 in the $B_5(\pi/2)$ basis and qubit 2 in the computational basis, the oracle applies the gate $(R_z(\pi/2) \otimes R_z(\pi/2)){\sf CZ}$ between $\ket{x_1}$ and $\ket{y}$~\cite{onewayexp2}, where ${\sf CZ}$ shifts the relative phase of the state $\ket{1}\ket{1}$ by $\pi$ with respect to the rest of the computational basis states. Here, $\ket{x_2}$ remains unaffected. By including the measurement of qubit 4 again, this gives the computation $\ket{x_1}\ket{y \oplus f(x)}\ket{x_2}={\sf CNOT}\otimes \openone \ket{+}\ket{-}\ket{+}\equiv [R_z(\pi/2) \otimes R_z(\pi/2) \otimes \openone][{\sf CZ} \otimes \openone][\openone \otimes {\sf H} \otimes \openone]\ket{+}\ket{-}\ket{+}$, up to local rotations $[R_z(-\pi/2)\otimes {\sf H}\,R_z(-\pi/2)\otimes \openone]_{631}$ incorporated into the FF stage. For BB(v), the roles of qubits 2 and 5 in the program for BB(iii) are interchanged, thus implementing a {\sf CNOT} between $\ket{x_2}$ and $\ket{y}$ instead. For BB(g) both qubits 2 and 5 are measured in the basis $B(\pi/2)$, thereby applying a {\sf CNOT} between $\ket{x_1}$ and $\ket{y}$, together with one between $\ket{x_2}$ and $\ket{y}$. Finally, for BB's (ii), (iv), (vi) and (viii) an additional $\sigma_x$ rotation is applied to qubit 3 at the FF stage using the measurement programs of BB's (i), (iii), (v) and (vii) respectively. The initial $\sigma_z$ rotation on qubit 4, along with the measurements and outcomes of qubits $1$, $3$ and $6$ constitute the algorithm, although only the query qubits need to be measured. The additions to the FF and measurements of qubits 2, 4 and 5 should be viewed as being carried out entirely by the oracle. 

In the refined version of the algorithm one replaces the ancilla $\ket{y}$ with an additional query qubit $\ket{x_3}$. The initial Hadamard gate is not required and qubit 4 can be removed, with qubit 3 as the input for $\ket{x_3}$. The resource requirement is therefore a five-qubit cluster state. Unfortunately, in this configuration it is not possible to implement all of the BB's that the oracle might perform for DJ$_3$. This point will be discussed in more detail later.   

{\it BV algorithm.-} In this algorithm~\cite{BV} an $n$-bit binary input $x \in \{0,1\}^{n}$ is used and the problem is to determine the hidden value $s \in \{0,1\}^{n}$ of the function $f_s(x)=s \cdot x$ carried out by an oracle. 
The oracle's transformation is $f_s(x): \{0,1\}^{n} \mapsto \{0,1\}$ and classically one needs to make $n$ queries in order to determine $s$~\cite{BV}. This problem can be seen as a more {\it sophisticated} database search compared to the one in Grover's search algorithm~\cite{Meyer, algoritmi}. Here, the oracle plays the role of a database where records are represented by vectors. Record $s$ is tagged and when the oracle is queried with a specific record $x$, it responds by providing information $s \cdot x$ signifying the similarity of the guess to the tagged record~\cite{Meyer}. This resembles the response from a web search-engine, where the pages are usually ordered by their relevance~\cite{Yuwono}. Using the BV algorithm only one query is necessary to find the tagged record. 
A similar method to the DJ algorithm is employed using the same input states. Taking the output state from the oracle $\ket{\psi_{out}}=2^{-n}\sum_{z}\sum_{x} (-1)^{x \cdot z + f_s(x)} \ket{z}\ket{-}\equiv2^{-n}\sum_{z}\sum_{x} (-1)^{s \cdot x}(-1)^{x \cdot z} \ket{z}\ket{-}$ we have, using the relation $2^{-n}\sum_{x}(-1)^{s \cdot x}(-1)^{x \cdot z}=\delta_{s,z}$, that $\ket{\psi_{out}}= \ket{s}\ket{-}$. Therefore measuring the output query register once reveals the value of $s$. As before, in the refined version, the ancilla qubit is removed and the oracle transformation is $\ket{x}\mapsto (-1)^{f_s(x)}\ket{x}$. The remainder of the procedure is performed as described above. 

BV$_1$ reduces to Deutsch's problem, thus we consider the fundamental version of the algorithm as BV$_2$. Here the required black boxes correspond to BB's (i), (v), (iii) and (vii) of DJ$_2$ for $s=00$, $01$, $10$ and $11$ respectively. Thus surprisingly the same cluster state $\ket{\Phi_{C}}$ can be used. For the refined version, BV$_3$, qubit 4 is again removed and qubit 3 becomes $\ket{x_3}$. However, it turns out that no two-qubit gates are required between the $\ket{x_i}$'s, only local $\sigma_z$ operations~\cite{Meyer}: a cluster state is not necessary for the refined version.
 \begin{figure}[t]
\psfig{figure=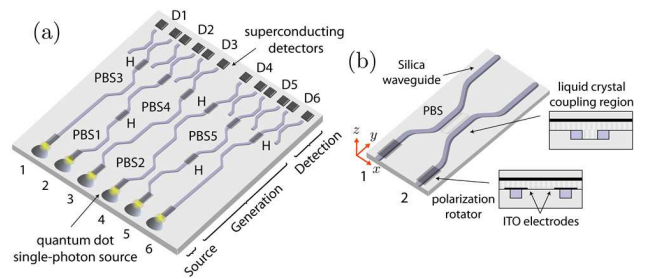,width=8.5cm}
\caption{Photonic on-chip setup for demonstrating the algorithms. {\bf (a)}:~The setup has three regions. (i) Quantum dot sources (1-6) produce single photons. (ii) A network of silica-on-silicon waveguide polarizing beamsplitters (PBSs 1-5) generate the cluster state via interference. (iii) Superconducting detectors detect the photons. The ${\sf H}$'s are Hadamard gates implemented by liquid crystal (LC) polarization rotators. {\bf (b)}:~PBS made of two waveguides with integrated LC. }
\label{setup}
\end{figure}

{\it On-chip photonic demonstration.-}
We now propose a setup that could be used to generate the cluster state $\ket{\Phi_C}$ and demonstrate the algorithms. Fig.~\ref{setup}~{\bf (a)} shows a network of photonic silica-on-silicon waveguides~\cite{Silica}, where the polarization degrees of freedom of a photon in mode $i$ embody a qubit with the association $\ket{0}_i \to \ket{H}_i$ and $\ket{1}_i \to \ket{V}_i$. Initially, photons are generated in waveguide modes 1 to 6 by optically pumping individual quantum dots~\cite{Shields} whose emission is directly coupled into the respective waveguide. The photons are then prepared in the state $\ket{+}$ by liquid crystal (LC) polarization rotators that are voltage-controlled by indium tin oxide (ITO) electrodes~\cite{chipPBS}. The action of the polarizing beamsplitters (PBS's) that follow is to apply a {\it fuse} operation $\ketbra{HH}{HH}+\ketbra{VV}{VV}$ on two photon modes when a single photon exits each output port~\cite{BrowneRudolph}. This occurs with probability $1/2$. The PBS component is depicted in more detail in Fig.~\ref{setup}~{\bf (b)}, where the region between two waveguides is filled with LC having no effect on $V$-polarized photons but switching those with $H$-polarization (see caption for further details)~\cite{chipPBS}. We note that such an on-chip component has not been considered before for quantum processing and could represent a versatile component in advanced photonic quantum technology~\cite{OBrien2}. Considering the action of the PBS's and polarization rotations ${\sf H}$ before the detection stage, it is straightforward to show that if one photon is present in each mode, leading to one photon at each detector-pair ${\sf Di}$ in coincidence, the state $\ket{\Phi_C}$ will be generated upon detection. All other possibilities do not lead to a six-photon coincidence and therefore can be post-selected out. For this purpose we consider superconducting single-photon detectors~\cite{Hadfield} embedded within the chip. The measurement programs are realized at the detector stage using the final polarization rotator and PBS before each detector-pair~\cite{onewayexp1,onewayexp2,onewayexp3}.   
\begin{figure}[t]
\psfig{figure=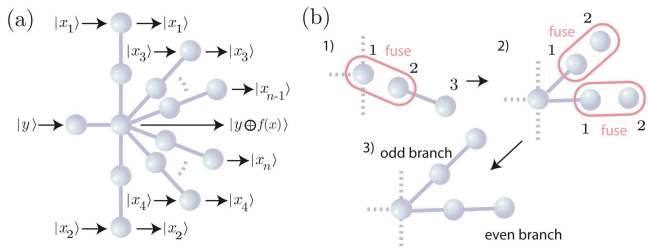,width=8.5cm}
\caption{Configuration for DJ$_n$ and BV$_n$. {\bf (a)}: Graph state depicting the in-out logical qubits. {\bf (b)}:~Construction by PBS fusion ($[\ketbra{HH}{HH}+\ketbra{VV}{VV}]_{12}$ followed by $\openone_1 \otimes {\sf H}_2$) from one pair of photons already fused in the state $(\openone \otimes {\sf H})1/\sqrt{2}[\ket{HH}+\ket{VV}]_{23}$ and two single photons in $\ket{+}$.}
\label{BVgen}
\end{figure}

{\it Scalability.-} We now investigate extending the method proposed here for performing arbitrary $n$-qubit versions of both algorithms. In order to be able to implement all BB functions for the standard (refined) DJ algorithm with $n \ge 3$, a Toffoli ({\sf C$^2$Z}) gate~\cite{DJ,MT} is required. A simple example is the balanced output $\{f(0),f(1),f(2),f(3),f(4),f(5),f(6),f(7)\}:= \{0,0,0,1,1,1,1,0 \}$ for which the standard (refined) version requires a Toffoli ({\sf C$^2$Z}) gate between $\ket{x_2}$, $\ket{x_3}$ and $\ket{y}$ ($\ket{x_1}$, $\ket{x_2}$ and $\ket{x_3}$), with $\ket{y}$ ($\ket{x_3}$) the target and a {\sf CNOT} ({\sf CZ}) gate between $\ket{x_1}$ and $\ket{y}$ ($\ket{x_3}$). Fortunately, it is always possible to implement a selection of constant and balanced BB's for arbitrary $n$ by using combinations of {\it controllable}-{\sf CNOT}'s~\cite{onewayexp2} between the query qubits $\ket{x_i}$ and the ancilla $\ket{y}$ ({\it controllable}-{\sf CZ}'s between the query qubits), as presented for DJ$_2$. This drastically reduces the resource requirements for demonstration purposes. For the BV algorithm one can see by inspection of the oracle's action $f_s(x)=s \cdot x=(x_1\wedge s_1)\oplus (x_2\wedge s_2) \oplus \cdots \oplus (x_n\wedge s_n)$ that it is easily generalized. In the standard (refined) case, the value of $\ket{x_i}$ determines whether or not to apply a $\sigma_x$ ($\sigma_z$) operation to $\ket{y}$ ($\ket{x_i}$). In Fig.~\ref{BVgen}~{\bf (a)} we show a graph state~\cite{Hein} that could be used for demonstrating both DJ$_n$ and BV$_n$. In Fig.~\ref{BVgen}~{\bf (b)} we show a possible way to scale up the resource via fuse operations, which could be realized using our proposed on-chip setup. Note that as each branch is created with probability 1/4, the success probability of {\it generating} the state decreases exponentially with $n$, however this has no effect on the performance of the algorithms. If needed, more economical generation techniques could be used for large $n$~\cite{BrowneRudolph}. With the use of on-chip photon-number resolving detectors~\cite{Divochiy} and appropriate network for error-correction one can achieve a scalable method to demonstrate the algorithms.

We have shown that fundamental versions of the DJ and BV algorithms can be performed on a six-qubit cluster state and investigated the minimal resources for scaling up to general $n$-qubit versions. A photonic on-chip setup was then proposed for realizing our scheme in a scalable context. The techniques described here could be applied equally well to other algorithms and protocols. Although we present an on-chip setting our proposal is readily applicable to other types of waveguide and bulk setups. As these settings require researchers from a wide range of the physical sciences working together, we expect this study to stimulate progress of practical and scalable QC in both a theoretical and experimental capacity.

We thank C. Di Franco and R. Prevedel for comments and acknowledge support from UK EPSRC and QIP IRC.

{\it Note added.-} After this work was written up we became aware of a related, although not as general, cluster state experiment using bulk optics~\cite{Vallonerel}. 


\end{document}